\documentclass[twocolumn, showpacs, preprintnumbers, superscriptaddress, aps, prl, floatfix, 10pt, tightenlines]{revtex4-1}  

\usepackage{CJKutf8}
\usepackage{latexsym}
\usepackage{amstext}
\usepackage{amsfonts}
\usepackage{dsfont}
\usepackage{color}
\usepackage{amssymb}
\usepackage{amsmath}
\usepackage{relsize}
\usepackage{amsxtra}
\usepackage{verbatim}
\usepackage{hyperref}
\usepackage{graphicx}
\usepackage{dcolumn}
\usepackage{subfigure}

\AtBeginDvi{\input{zhwinfonts}}

\begin{document}
\begin{CJK*}{UTF8}{zhkai}

\title{A second-order spin-flop transition in collinear two-sublattice antiferromagnets}

\author{Haifeng Li} 
\email{h.li@fz-juelich.de}
\affiliation{J$\ddot{u}$lich Centre for Neutron Science JCNS, Forschungszentrum J$\ddot{u}$lich GmbH, Outstation at Institut Laue-Langevin, Bo$\hat{\imath}$te Postale 156, F-38042 Grenoble Cedex 9, France}
\affiliation{Institut f$\ddot{u}$r Kristallographie der RWTH Aachen University, D-52056 Aachen, Germany}

\date{\today}

\begin{abstract}
Identifying the nature of a spin-flop (SFO) transition, first- or second-order (FO or SO), remains a major challenge in condensed-matter physics due to the technically undistinguishable effect of misalignment between applied-field direction and the relevant antiferromagnetic (AFM) easy axis. A classical SFO transition is believed to be of FO in character. Here a mean-field theoretical calculation endowed with AFM exchange interaction (\emph{J}), easy axis anisotropy ($\gamma$), uniaxial single-ion exchange anisotropy (\emph{D}), and Zeeman coupling to a magnetic field parallel to the easy axis unambiguously reveals that a SO SFO transition indeed exists by virtue of its relatively lower free energy. Their equilibrium phase conditions are found to be: $D \geq 0$ (FO); $-\frac{1}{2} \gamma < D < 0$ (SO). Compared numerically to the associated AFM and spin-flip phases, the deduced SO SFO transition results from a negative single-ion anisotropy which is restricted to a certain range by the anisotropic exchange interaction.
\end{abstract}

\pacs{02.10.{\texttt{-}}v, 64.10.{\texttt{+}}h, 75.10.Hk, 75.30.{\texttt{-}}m}

\maketitle
\end{CJK*}

\begin{figure*}
\centering \includegraphics[width = 0.856\textwidth] {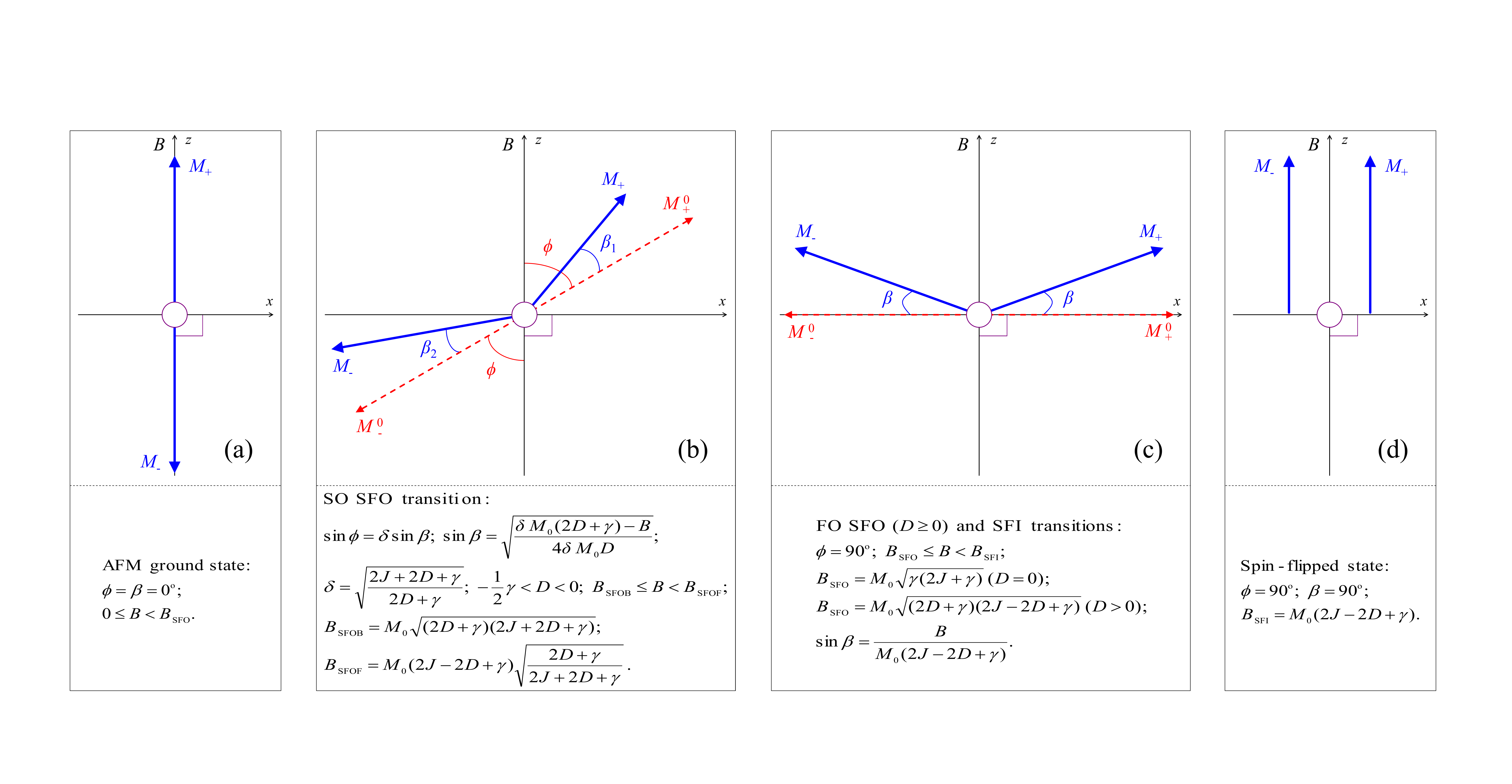}
\caption{(color online)
Schematic SFO and SFI transitions for a collinear two-sublattice antiferromagnet.
(a) In a normal AFM state, the AFM easy axis $M^0_-M^0_+$ coincides with localized sublattice moments $M_+$ and $M_-$, and all are supposed to be parallel to the $z$ direction when $0 \leq B < B_{\texttt{SFOB}}$. Here $\phi = \beta_1 = \beta_2 = 0^{\circ}$. (b) A SO SFO transition in the range of fields $B_{\texttt{SFOB}} \leq B < B_{\texttt{SFOF}}$. $\phi$ denotes an angle of the AFM easy direction away from the $z$ axis. $\beta_1$ and $\beta_2$ correspond to angles of sublattice moments $M_+$ and $M_-$ away from the $M^0_-M^0_+$ axis, respectively. Here $0 < \phi < 90^{\circ}$, and $\beta_1 \equiv \beta_2 = \beta$ in the saturation magnetic state at sufficiently low temperatures. (c) When $\phi = 90^{\circ}$, sublattice moments are rightly flopped at $B_{\texttt{SFOF}}$ and then tilted away from the $x$ axis by an angle $\beta$. Here $\beta \in (0^{\circ}, 90^{\circ})$ and $B_{\texttt{SFOF}} \leq B < B_{\texttt{SFI}}$ in the process of a SFI transition. (d) Sublattice moments $M_+$ and $M_-$ are completely aligned along the \emph{\textbf{B}} (i.e. $z$) direction in a strong enough field $B_{\texttt{SFI}}$ so that $\beta = 90^{\circ}$. The corresponding equilibrium phase conditions of the deduced magnetic states are listed in the low panels of (a-d), respectively.
}
\label{AllP}
\end{figure*}

For a collinear antiferromagnet below the N\'{e}el temperature, when magnetic field (\emph{\textbf{B}}) applied along its antiferromagnetic (AFM) easy axis reaches a critical point $(B_{\texttt{SFO}})$, the AFM sublattice spins suddenly rotate 90$^\circ$ so that they will be perpendicular to the original AFM easy axis. This is the traditional spin-flop (SFO) transition, typically a first-order (FO) type. After this, the flopped spins are gradually tilted along the field direction with increasing field strength $(B > B_{\texttt{SFO}})$ until they are completely aligned at a sufficiently high field $(B_{\texttt{SFI}})$, which is the so-called spin-flip (SFI) transition. These magnetic phase transitions with field are schematically sketched in Fig.~\ref{AllP}.

N\'{e}el for the first time proposed theoretically the possibility for a SFO transition in 1936 \cite{Neel1936}. Subsequently, it was observed experimentally in a CuCl$_2\cdot$2H$_2$O single crystal \cite{Poulis1951}. Since then, the SFO transition has been extensively investigated, and the corresponding phenomenological theory has been comprehensively developed, generally confirming that it is of FO in nature \cite{Ranicar1967, Rohrer1975, Butera1981, Filho1991}. However, most of the reported sharp SFO transitions \cite{Oliveira1978, Filho1991, Becerra1993} display no magnetic hysteresis effect characteristic of a FO phase transition. This was attributed either to a low magnetic anisotropy \cite{Filho1991, Becerra1993} or to a softening of surface magnons \cite{Keffer1973}. In addition, some FO SFO transitions are obviously continuous occurring in a broad field range, which was ascribed either to a domain effect resulting from the inhomogeneous character of the diluted systems or to a misalignment of the applied field with regard to the AFM easy axis \cite{Rohrer1975, King1979}. On the other hand, this kind of continuous magnetic behavior, the absence of the magnetic hysteresis, and the experimental observation of a possible intermediate phase in the SFO compound CoBr$_2\cdot6$[0.48D$_2$O, 0.52H$_2$O] \cite{Smeets1982} cast considerable doubt on the nature of SFO transition and in addition may indicate a second-order (SO) type phase transition. Indeed, early theoretical calculations \cite{Yamashita1972, Becerra1974} predicted an intermediate regime bordering with the AFM and the spin-flopped states. However, this has not yet been confirmed based on the principle of minimum total potential energy. In addition, a deviated (which is called \texttt{"}freely-rotating\texttt{"} in this study) ferromagnetic (FM) phase \cite{Yamashita1972, Prystasz1982} was also predicted, but has never been experimentally observed so far.

\begin{figure}
\centering \includegraphics[width = 0.4718\textwidth] {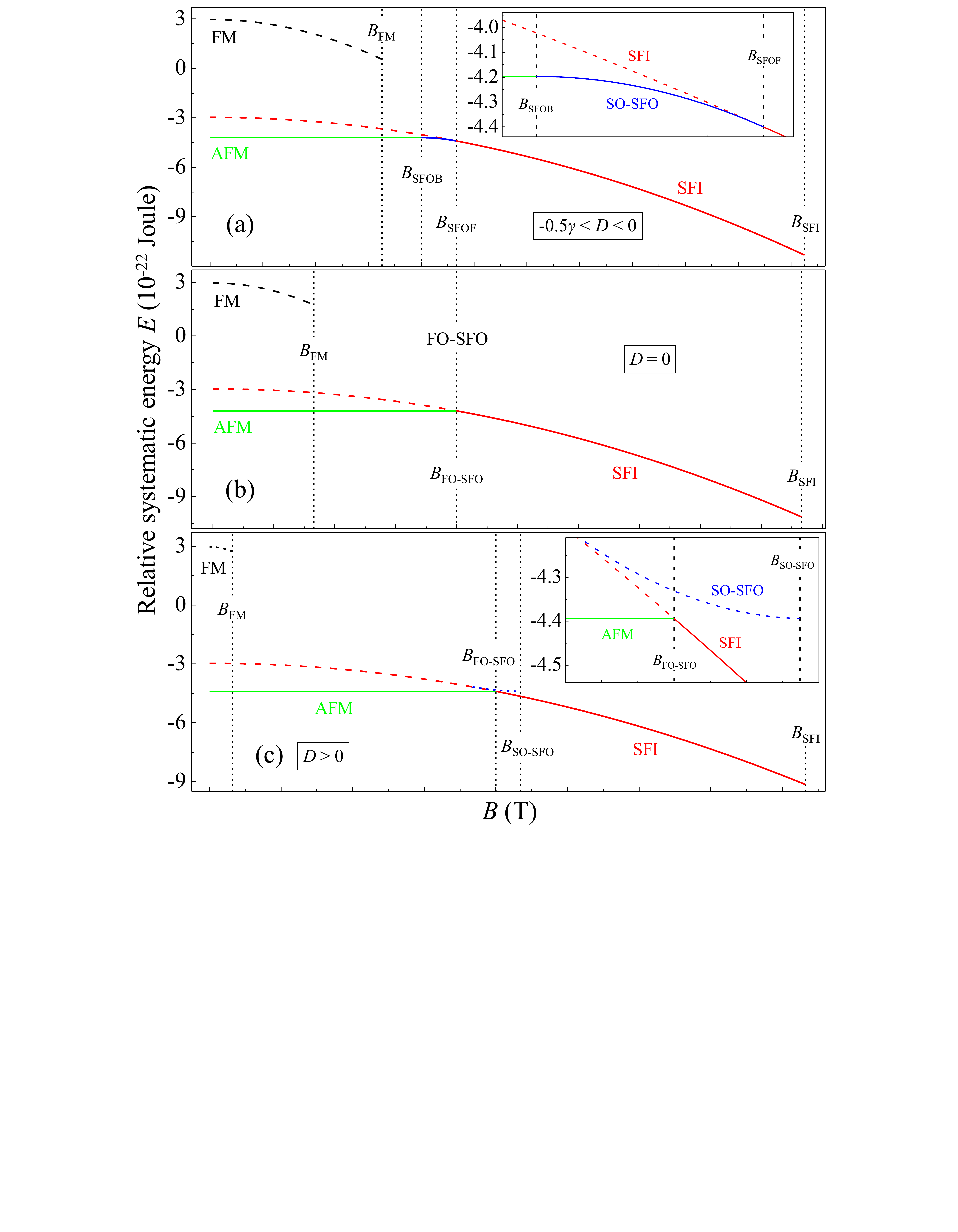}
\caption{(color online)
Calculated relative systematic energies of the deduced magnetic states.
(a) When $-\frac{1}{2}\gamma < D < 0$, a SO SFO transition occurs in the field range from $B_{\texttt{SFOB}}$ to $B_{\texttt{SFOF}}$. (b) When $D = 0$, a FO SFO transition happens at $B_{\texttt{FO-SFO}}$ = $B_{\texttt{SFOB}}$ = $B_{\texttt{SFOF}}$ = $M_0\sqrt{\gamma (2J + \gamma)}$. (c) When $D > 0$, a FO SFO transition occurs at $B_{\texttt{FO-SFO}}$ = $M_0\sqrt{(2D + \gamma)(2J - 2D + \gamma)}$. In (a-c), the calculated energies of the AFM state, in the process of the SFI transition, and during the free rotation of the FM-like state while $B_{\texttt{FM}} \leq M_0(\gamma - 2D)$ are all plotted for a clear comparison. Insets of (a) and (c) show an enlargement of the most interesting field regimes. The $E_{\texttt{SFI}}$ under $B < B_{\texttt{SFOF}}$ is also shown (dashed read line). Indeed, it is higher than those of the AFM state and the state during the process of the SFO transition. In (c), the mathematically permissible \emph{E} of the SO SFO transition is also displayed. In any case, the solid lines as shown in (a-c) represent the theoretically-allowed magnetic ground states with field \emph{\textbf{B}}.
}
\label{AllE}
\end{figure}

\begin{figure}
\centering \includegraphics[width = 0.48\textwidth] {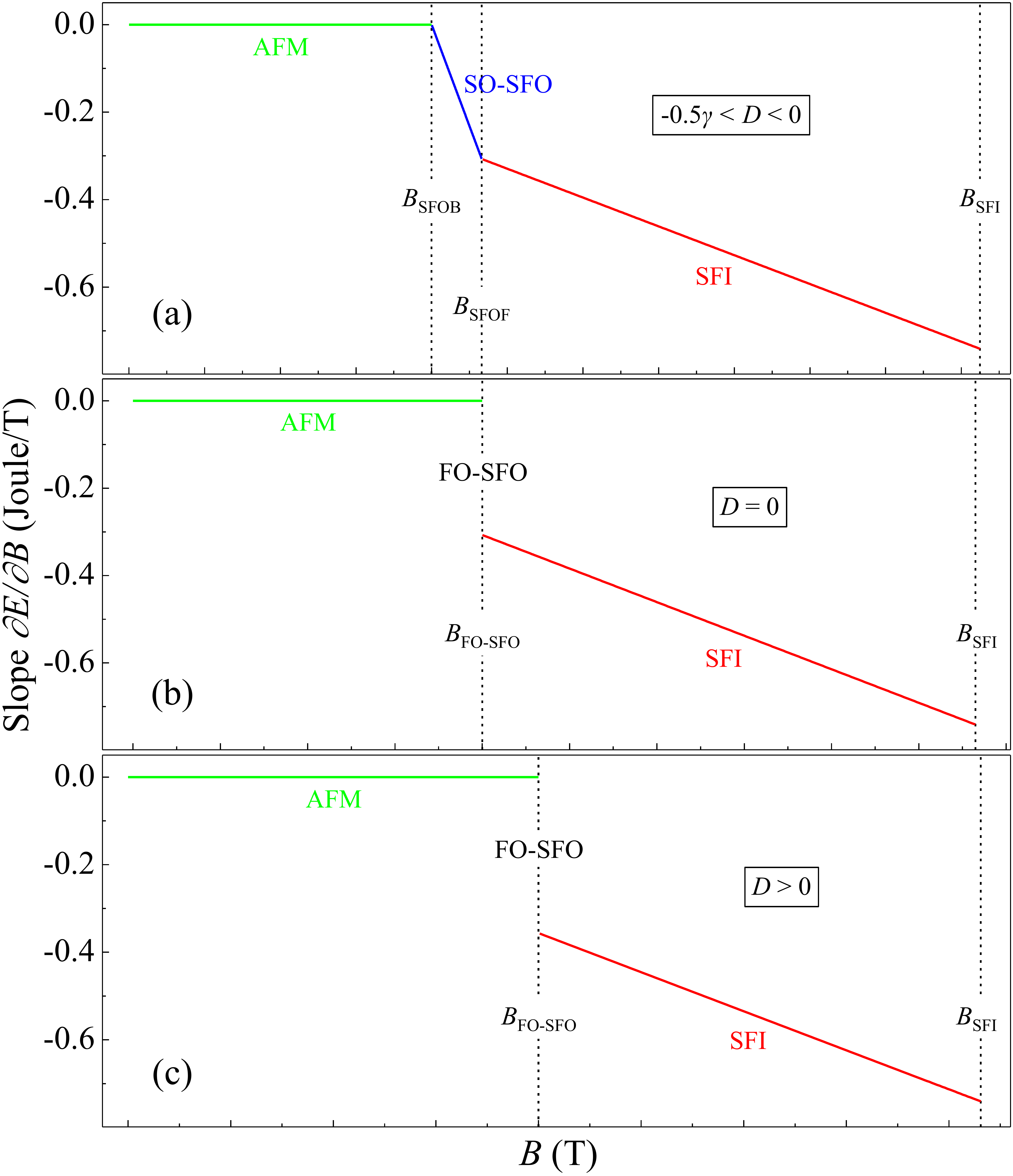}
\caption{(color online)
Variation of the energy slope $(\partial E / \partial B)$ as a function of magnetic field, corresponding to Fig.~\ref{AllE}.
(a) When $-\frac{1}{2}\gamma < D < 0$, the first derivative of the systematic energy \emph{E} with regard to field \emph{B} equals to zero in the AFM state and displays a continuous change while undergoing the SO SFO transition from $B_{\texttt{SFOB}}$ to $B_{\texttt{SFOF}}$ and then the SFI transition from $B_{\texttt{SFOF}}$ to $B_{\texttt{SFI}}$. (b) and (c) When $D = 0$ (b) and $D > 0$ (c), an abrupt change in the slope occurs at the FO SFO transition field $B_{\texttt{FO-SFO}}$.
}
\label{slope}
\end{figure}

\begin{figure}
\centering \includegraphics[width = 0.48\textwidth] {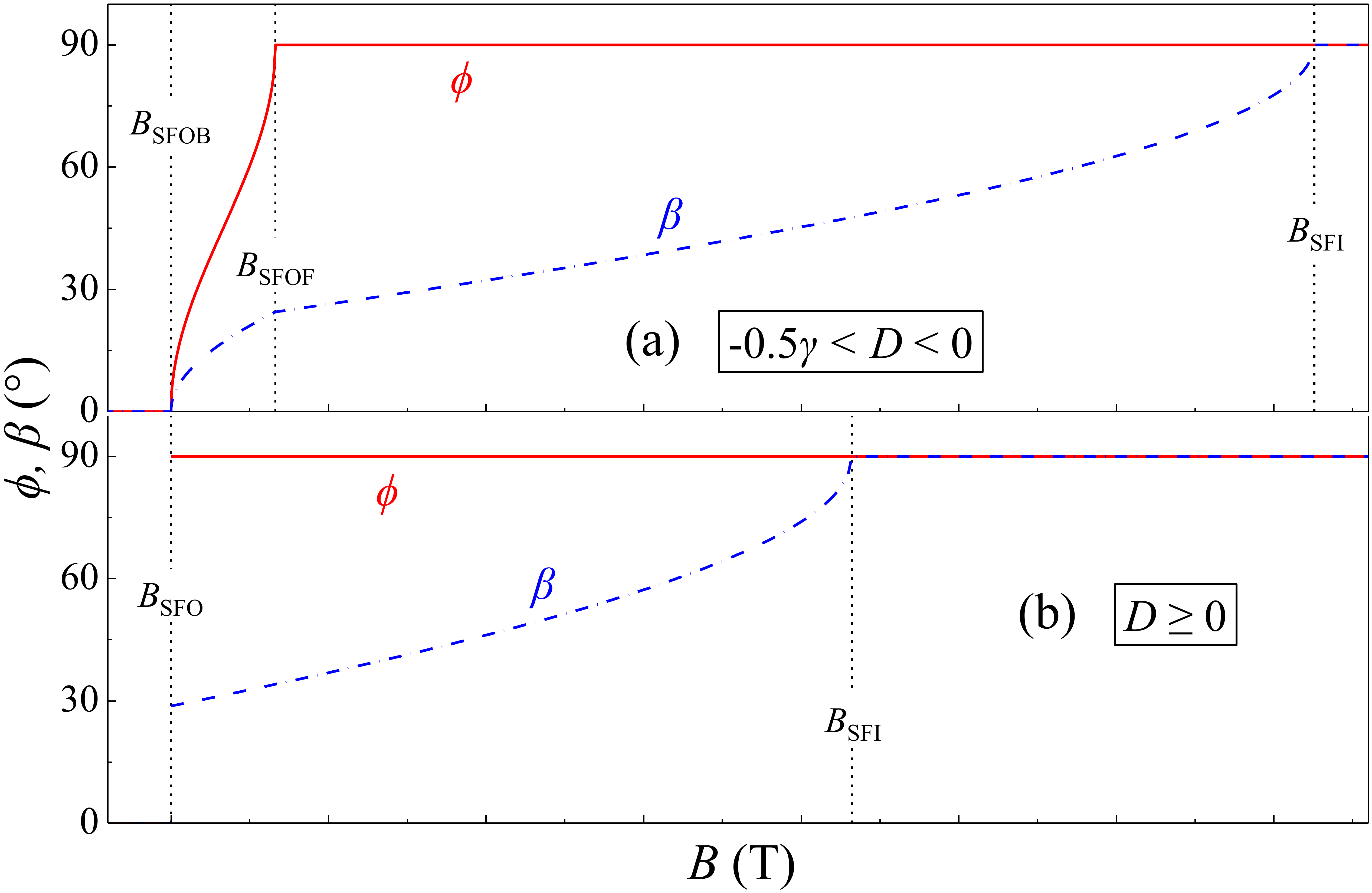}
\caption{(color online)
Variations of the angles $\phi$ and $\beta$ with magnetic filed \emph{\textbf{B}}.
(a) When $-\frac{1}{2}\gamma < D < 0$, $\phi$ and $\beta$ increase continuously in the range of fields $B_{\texttt{SFOB}} \leq B < B_{\texttt{SFOF}}$, suggesting a SO SFO transition. When $B_{\texttt{SFOF}} \leq B \leq B_{\texttt{SFI}}$, $\phi = 90^\circ$ and $\beta$ keeps the continuous growth up to 90$^\circ$ at $B_{\texttt{SFI}}$. (b) When $D \geq 0$, $\phi$ suddenly becomes 90$^\circ$ at $B_{\texttt{SFO}} = B_{\texttt{SFOB}} = B_{\texttt{SFOF}}$ indicative of a FO SFO transition, after which the magnetic state enters into the process of a SFI transition. In (a) ((b)), below $B_{\texttt{SFOB}}$ ($B_{\texttt{SFO}}$), $\phi = \beta = 0^\circ$; above $B_{\texttt{SFI}}$, $\phi = \beta = 90^\circ$.
}
\label{betaphi}
\end{figure}

In this paper, the magnetic-field induced SFO and SFI phase transitions of the localized collinear antiferromagnets are explored via a mean-field theoretical calculation, which conclusively rules out the possibility for a deviated FM-like state \cite{Yamashita1972} and unambiguously reveals that a SO SFO transition indeed exists theoretically through comparing for the first time their relative systematic energies.

\textbf{Methods}

The calculation performed here is limited to the purely-localized collinear AFM systems, ignoring the effect of valence electrons on magnetic couplings. In case of considering a two-sublattice spin configuration (Fig.~\ref{AllP}), the corresponding Hamiltonian terms of such kind of systems considered in a magnetic field \emph{\textbf{B}} consist principally of magnetic exchange, spin-exchange anisotropy, single-ion anisotropy, and Zeeman coupling. Assuming that an AFM easy direction with localized sublattice moments $M_+$ and $M_-$ is along the \emph{z} axis (Fig.~\ref{AllP}(a)) and that the completely-flopped spins are parallel to the \emph{x} axis (Figs.~\ref{AllP}(b) and (c)), the sublattice-moment vectors within the \emph{xz} plane (Fig.~\ref{AllP}(b)) can thus be written as:
\setlength\arraycolsep{1.4pt} 
\begin{eqnarray}
\label{MVectors}
\begin{split}
\widehat{M}_+ &= \text{\textcolor[rgb]{1.00,1.00,1.00}{-a}} M_+ [\hat{x} sin(\phi - \beta_1) + \hat{z} cos(\phi - \beta_1)] \text{\textcolor[rgb]{1.00,1.00,1.00}{-} and}  \\
\widehat{M}_- &= -M_- [\hat{x} sin(\phi + \beta_2) + \hat{z} cos(\phi + \beta_2)],
\end{split}
\end{eqnarray}
respectively, where $\hat{x}$ and $\hat{z}$ are the unit-vectors along the $x$ and $z$ axes, respectively. Therefore, the resultant systematic energy (\emph{E}) within the mean-field approximation can be calculated by:
\begin{eqnarray}
\label{energy1}
\begin{split}
\emph{E} = & \emph{J} \emph{M}_+ \cdot \emph{M}_- + \gamma \emph{M}_+^z \emph{M}_-^z - D[(\emph{M}_+^z)^2 + (\emph{M}_-^z)^2]     \\
&- B (\emph{M}_+^z + \emph{M}_-^z)                                                                                                \\
= & -\emph{J} M_+M_- \cos(\beta_1 + \beta_2)                                                                                      \\
&- \gamma M_+M_- \cos(\phi - \beta_1) \cos(\phi + \beta_2)                                                                        \\
&- D [M^2_+ \cos^2(\phi - \beta_1) + M^2_- \cos^2(\phi + \beta_2)]                                                                \\
&- B [M_+ \cos(\phi - \beta_1) - M_- \cos(\phi + \beta_2)],
\end{split}
\end{eqnarray}
where the four terms in turn denote the four Hamiltonian components as the foregoing remarks, and \emph{J}, $\gamma$, and \emph{D} are the universal magnetic-coupling, anisotropic-exchange, and single-ion anisotropic constants, respectively. At a nonzero temperature point, with increasing magnetic field \emph{\textbf{B}} $(\parallel z$ axis) as shown in Figs.~\ref{AllP}(a) and (b), sublattice moment $M_+$ ($M_-$) increases (decreases) as a consequence, which leads to $\beta_1 < \beta_2$. In the saturation magnetic state, i.e., $M_+ \equiv M_- = M_0$, at low enough temperatures, $\beta_1 \equiv \beta_2 = \beta$. Hence, Eq.~(\ref{energy1}) can be simplified as:
\begin{eqnarray}
\label{energy2}
\begin{split}
\emph{E} =& -\emph{J} M_0^2 \cos(2\beta) - \gamma M_0^2 \cos(\phi - \beta) \cos(\phi + \beta)                                     \\
&- D M_0^2[\cos^2(\phi - \beta) + \cos^2(\phi + \beta)]                                                                           \\
&- B M_0[\cos(\phi - \beta) - \cos(\phi + \beta)]                                                                                 \\
= & -\emph{J} M_0^2 \cos(2\beta) - \frac{\gamma M_0^2}{2} [\cos(2\phi) + \cos(2\beta)]                                            \\
&- D M_0^2[1 + \cos(2\phi) \cos(2\beta)]                                                                                          \\
&- 2 B M_0 \sin \phi \sin \beta.
\end{split}
\end{eqnarray}

\textbf{Results}

Possible magnetic equilibrium states can be derived from different combinations of the FO partial differential equations, i.e., $\frac{\partial E}{\partial \beta} = \frac{\partial E}{\partial \phi} =$ 0 with Eq.~(\ref{energy2}), as well as the corresponding boundary conditions:
\begin{eqnarray}
\label{AFM-SFO}
&\left\{
\begin{array} {l l}
2 D M_0 \sin \phi \cos (2 \beta) + \gamma M_0 \sin \phi - B \sin \beta = 0, \\
2 J M_0 \sin \beta + 2 D M_0 \cos (2 \phi) \sin \beta + \\
\gamma M_0 \sin \beta - B \sin \phi = 0;
\end{array}
\right. \\
\label{SFO}
&\left\{
\begin{array} {l l}
\cos \phi = 0, \quad \quad \quad \quad \quad \quad \quad \quad \quad \quad \quad \quad \quad \quad \quad \quad \quad \\
2 J M_0 \sin \beta + 2 D M_0 \cos (2 \phi) \sin \beta + \\
\gamma M_0 \sin \beta - B \sin \phi = 0;
\end{array}
\right. \\
\label{FM}
&\left\{
\begin{array} {l l}
2 D M_0 \sin \phi \cos (2 \beta) + \gamma M_0 \sin \phi - B \sin \beta = 0, \\
\cos \beta = 0;
\end{array}
\right. \\
\label{SFI}
&\left\{
\begin{array} {l l}
\cos \phi = 0, \quad \quad \quad \quad \quad \quad \quad \quad \quad \quad \quad \quad \quad \quad \quad \quad \quad \\
\cos \beta = 0.
\end{array}
\right.
\end{eqnarray}

In the following, the four combinations~(\ref{AFM-SFO}-\ref{SFI}) will tentatively be solved: (i) Firstly, the combination~(\ref{AFM-SFO}) involves the most formidable challenge, and one can obtain ultimately two solutions: (A) $\sin \phi = \sin \beta = 0$, i.e., $\phi = \beta = 0$; (B) $\sin \phi = \delta \sin \beta$, where $\sin \beta = \sqrt{\frac{\delta M_0 (2D + \gamma) - B}{4 \delta M_0 D}}$ and $\delta = \sqrt{\frac{2 J + 2 D + \gamma}{2D + \gamma}}$. The former case (A) is associated with an AFM ground state as shown in Fig.~\ref{AllP}(a), while the latter case (B) signifies a correlated change of $\phi$ with $\beta$. (ii) The combination~(\ref{SFO}) implies that $\phi = \frac{\pi}{2}$, and $\sin \beta = \frac{B}{M_0(2J - 2D + \gamma)}$, which corresponds to the process of a SFI transition (Fig.~\ref{AllP}(c)). When $B = B_{\texttt{SFI}} = M_0(2J - 2D + \gamma)$, $\beta = \frac{\pi}{2}$, implying a spin-flipped state (Fig.~\ref{AllP}(d)). Therefore, the SFI transition field $B_{\texttt{SFI}}$ depends not only on the moment size $M_0$ but also on the values of $J, \gamma$, and $D$. (iii) From the combination~(\ref{FM}), one can deduce that $\beta = \frac{\pi}{2}$ which is independent of $\phi$, and $\sin \phi = \frac{B}{M_0(\gamma - 2D)}$. When $\beta = \frac{\pi}{2}$, both sublattice moments $M_+$ and $M_-$ are perpendicular to the AFM axis $M^0_-M^0_+$, forming a freely-rotating FM-like state. The value of $\phi$ can intrinsically be modified by a change in magnetic field $B$. (iv) The simplest combination~(\ref{SFI}) indicates that $\phi = \beta = \frac{\pi}{2}$, which corresponds to a spin-flipped state as schematically shown in Fig.~\ref{AllP}(d).

\textbf{Discussion}

As shown in Fig.~\ref{AllP}, $0^\circ \leq \phi \leq 90^\circ$. Consequently, there are two boundary magnetic fields corresponding to the SFO transition (i.e. the second solution of the combination~(\ref{AFM-SFO})). When $\phi = 0$, $\sin \phi = \delta \sin \beta = \delta \sqrt{\frac{\delta M_0 (2D + \gamma) - B}{4 \delta M_0 D}} = 0$. One can deduce that the initial magnetic field for the beginning of the SFO transition is $B_{\texttt{SFOB}} = M_0 \sqrt{(2D + \gamma)(2J + 2D + \gamma)}$. When $\phi = \frac{\pi}{2}$, $\sin \phi = \delta \sqrt{\frac{\delta M_0 (2D + \gamma) - B}{4 \delta M_0 D}} = 1$, therefore, the final magnetic field for the ending of the SFO transition is $B_{\texttt{SFOF}} = M_0 (2J - 2D + \gamma) \sqrt{\frac{2D + \gamma}{2J + 2D + \gamma}}$. When $B_{\texttt{SFOB}} \geq B_{\texttt{SFOF}}$, one can derive the precondition of a FO SFO transition: $D \geq 0$. On the other hand, when $B_{\texttt{SFOB}} < B_{\texttt{SFOF}}$, i.e., $-\frac{1}{2} \gamma < D < 0$, a surprising SO SFO transition occurs spontaneously, which originates from a negative single-ion anisotropy (relative to the magnetic interaction) as well as its competition with the anisotropic-exchange interaction.

To calculate the energy scales of the deduced magnetic states from the four combinations~(\ref{AFM-SFO}-\ref{SFI}), one can substitute their respective equilibrium conditions as discussed above back into Eq.~\ref{energy2} and then obtain:
\begin{align}
\label{AFME}
&E_{\texttt{AFM}} = -(J + \gamma + 2 D) M_0^2 \text{\textcolor[rgb]{1.00,1.00,1.00}{a}}(0 \leq B < B_{\texttt{SFO}});                                   \\
\label{SFIE}
&E_{\texttt{SFI}} = -J M_0^2 - \frac{B^2}{2J - 2D + \gamma} \text{\textcolor[rgb]{1.00,1.00,1.00}{a}}(B_{\texttt{SFO}} \leq B < B_{\texttt{SFI}});      \\
\label{SFIDE}
&E_{\texttt{SFID}} = (-3J + 2 D - \gamma) M_0^2 \text{\textcolor[rgb]{1.00,1.00,1.00}{a}}(B = B_{\texttt{SFI}});                                        \\
\label{FMlikeE}
&E_{\texttt{FM-like}} = J M_0^2 - \frac{B^2}{\gamma - 2D} \text{\textcolor[rgb]{1.00,1.00,1.00}{a}}(0 \leq B \leq M_0 (\gamma -2 D));
\end{align}
the one corresponding to the deduced SO SFO transition (Fig.~\ref{AllP}(b)) is presented separately due to its complexity:
\begin{widetext}
\begin{align}
\label{SOSFOE}
E_{\texttt{SO-SFO}} = &-JM_0^2(1 - 2\sin^2\beta) - \gamma M^2_0 (1 - \sin^2\beta - \delta^2\sin^2\beta) - DM^2_0 [1 + (1 - 2\delta^2\sin^2\beta)(1 - 2\sin^2\beta)] - 2BM_0\delta\sin^2\beta \nonumber \\
&\text{\textcolor[rgb]{1.00,1.00,1.00}{a}}(\sin \beta = \sqrt{\frac{\delta M_0 (2D + \gamma) - B}{4 \delta M_0 D}}; \delta = \sqrt{\frac{2 J + 2 D + \gamma}{2D + \gamma}}; -\frac{1}{2}\gamma < D < 0; B_{\texttt{SFOB}} \leq B < B_{\texttt{SFOF}}).
\end{align}
\end{widetext}

To quantitatively compare the energies (Eqs.~(\ref{AFME}-\ref{SOSFOE})), in the following the comparison will be divided into three parts based on the value of $D$. Firstly, the case of the SO SFO transition under the condition of $-\frac{1}{2}\gamma < D <0$ is presented. Supposing that $M_\texttt{0}$ = 4 $\mu_\texttt{B}$, \emph{J} = 2 T/$\mu_\texttt{B}$, \emph{D} = -0.2 T/$\mu_\texttt{B}$, and $B_{\texttt{SFOB}} = 8$ T, which are substituted into $B_{\texttt{SFOB}} = M_0 \sqrt{(2D + \gamma)(2J + 2D + \gamma)}$, one thereby gets $\gamma \sim$ 1.228 T/$\mu_\texttt{B}$ which satisfies the boundary condition $D > -\frac{1}{2}\gamma$. Based on these values, one can get that $B_{\texttt{SFOF}} = M_0 (2J - 2D + \gamma) \sqrt{\frac{2D + \gamma}{2J + 2D + \gamma}} \sim 9.325$ T, $\delta = \sqrt{\frac{2 J + 2 D + \gamma}{2D + \gamma}} \sim 2.414$, and $B_{\texttt{SFI}} = M_0 (2J - 2D + \gamma) \sim 22.514$ T. Therefore, the relative systematic energies of all possible magnetic states can be calculated as shown in Fig.~\ref{AllE}(a). Secondly, in the above assumed parameters if one sets $D = 0$ T/$\mu_\texttt{B}$, then $\gamma \sim$ 0.828 T/$\mu_\texttt{B}$, and $B_{\texttt{SFOF}} = B_{\texttt{SFOB}} = 8$ T, which corresponds to the FO SFO transition. The calculated relative energies at $D = 0$ T/$\mu_\texttt{B}$ are shown in Fig.~\ref{AllE}(b). Thirdly, in the case of $D > 0$, $B_{\texttt{SFOB}} > B_{\texttt{SFOF}}$. To extract the exact field for the FO SFO transition, solving $E_{\texttt{AFM}}$ (Eq.~\ref{AFME}) = $E_{\texttt{SFI}}$ (Eq.~\ref{SFIE}) yields that $B_{\texttt{SFO}}$ = $M_0\sqrt{(2D + \gamma)(2J - 2D + \gamma)}$. If one sets $D = 0.2$ T/$\mu_\texttt{B}$, then $\gamma \sim$ 0.561 T/$\mu_\texttt{B}$, and $B_{\texttt{SFI}} \sim 16.645$ T. The corresponding energy scales at $D = 0.2$ T/$\mu_\texttt{B}$ are displayed in Fig.~\ref{AllE}(c). It is clear that in the field range of $B \leq B_{\texttt{FM}}$, the relative systematic energy $E_{\texttt{FM}}$ is always higher than those of other allowed magnetic states (Fig.~\ref{AllE}), indicating that the freely-rotating FM-like state doesn't exist at all. As shown in Fig.~\ref{AllE}(a), an AFM state persists up to $B_{\texttt{SFOB}}$, then a SO SFO transition occurs in the range of fields $B_{\texttt{SFOB}} \leq B \leq B_{\texttt{SFOF}}$, followed by a SFI transition at $B > B_{\texttt{SFOF}}$. Finally, all sublattice spins are aligned along the field direction at $B_{\texttt{SFI}}$. By contrast, as shown in Figs.~\ref{AllE}(b) and (c), an antiferromagnet experiences a FO SFO transition at $B_{\texttt{FO-SFO}}$ and then enters directly into the process of a SFI transition.

The nature of a SFO transition can also be recognized by the character, continuous or discontinuous, of the first derivative of the energy (Fig.~\ref{AllE}) with regard to magnetic field based on the Ehrenfest's criterion \cite{Tari2003} for the FO and SO phase transitions. A continuous slope change is clearly illustrated in Fig.~\ref{slope}(a) where one can easily deduce that the second derivative $\partial ^2 E / \partial ^2 B$ is indeed discontinuous, whereas an abrupt change in the slope is obviously displayed at $B_{\texttt{FO-SFO}}$ in Figs.~\ref{slope}(b) and (c). To better understand the magnetic phase transitions with field, the values of the angles $\phi$ and $\beta$ (Fig.~\ref{AllP}) for all deduced magnetic states are also calculated in the whole field range as shown in Fig.~\ref{betaphi}. The SO SFO transition (Fig.~\ref{betaphi}(a)) and the FO SFO transition (Fig.~\ref{betaphi}(b)) are especially clear in terms of the variation of $\phi$ and $\beta$ with field. Until now, it can firmly be concluded that a SO SFO transition indeed exists theoretically.

When $\phi = 90^{\circ}$ and $\beta = 0^{\circ}$, the systematic energy of the classical SFO transition with a 90$^{\circ}$ rotation of the AFM spins is calculated as $E_{\texttt{C-FO-SFO}} = -J M_0^2$ according to Eq.~\ref{energy2}. By comparing $E_{\texttt{C-FO-SFO}}$ with that of the AFM state, i.e. $E_{\texttt{AFM}}$ (Eq.~\ref{AFME}), it is clear that the classical SFO transition occurs only when $D = \gamma = 0$.

In summary, a consistent mean-field calculation of the SFO and SFI transitions has been performed for the localized collinear antiferromagnets. In this study, two special magnetic states with field are derived: a freely-rotating FM-like state and a SO SFO transition. Based on the quantitative comparison of the ground-state energies, the former case has been clearly ruled out. But the latter case indeed exists, which is a sharp contrast to classical theories where the traditional SFO transition displays a FO fashion. This model calculation unifies AFM state, FO and SO SFO transitions, spin-flopped state, SFI transition as well as spin-flipped state. Their respective boundary conditions are extracted and listed in Fig.~\ref{AllP}. Inelastic neutron scattering on suitable real SFO compounds to extract the relevant parameters for a verification of the deduced boundary conditions would be of great interest, and Eq.~\ref{energy2} merits a tentative expansion with more agents.



\end{document}